\documentclass[aps,prl,twocolumn,showpacs]{revtex4}
\usepackage[utf8]{inputenc}
\usepackage{amsmath}
\usepackage{amssymb}
\usepackage{amsfonts}

\newcommand{\ie}{\textsl{i.e.}}
\newcommand{\eg}{\textsl{e.g.}}

\newcommand{\Eg}{\textsl{E.g.}}
\newcommand{\upd}{\mathrm{d}}

\begin{document}

\title{Generalizing Planck's law: \\Nonequilibrium emission of excited media}

\author{K.~Henneberger}
\email{henneb@physik3.uni-rostock.de}
\affiliation{University of Rostock, Institute of Physics, 18051 Rostock, Germany}

\pacs{42.50.Ct, 42.50.Nn, 78.45.+h} 

\begin{abstract}
Using a quantum-kinetic many-body approach, exact results for the interacting 
system of field and matter in a specified geometry are presented.
It is shown that both the spectral function of photons 
and the field fluctuations split up into vacuum- and medium-induced 
contributions, for which explicit expressions are derived. Using 
Poynting's theorem, the incoherent emission is analyzed and related to the coherent 
absorption as may be measured in a linear transmission-reflection experiment. 
Their ratio defines the medium-induced population of the modes of the transverse 
electromagnetic field and so generalizes Planck's law to an arbitrarily 
absorbing and dispersing medium in a nonequilibrium steady state. 
For quasi-equilibrium, this population develops into a Bose distribution 
whose chemical potential marks the crossover from absorption to gain and, 
also, characterizes the degree of excitation. Macroscopic quantum 
phenomena such as lasing and quantum condensation are discussed on this footing.
\end{abstract}

\maketitle

\section{Introduction}
Since the establishment of Kirchhoff's law till today, emission and absorption
are the fundamental observables in spectroscopic investigations. Assuming 
both ideal thermal radiation outside and thermodynamic equilibrium inside a 
medium and, also, between outside and inside, Kirchhoff's law 
$e(\omega)= n(\omega,T)a(\omega)$ fixes the ratio of emission $e$ to 
absorption $a$ as a universal function $n$ of frequency $\omega$ and 
temperature $T$. Thus, emission can be traced back to (i) absorption being 
observable in transmission-reflection experiments using classical 
(coherent) light, and (ii) the spectral distribution $n$ of thermal light,
which later was found to be a Bose function by Planck. However, if the very 
restricting assumption of complete thermodynamical equilibrium is removed, 
the emission of a medium is to be considered as a pure quantum effect, and 
quantum-optical and quantum-kinetic methods have to be used to describe the 
interacting many-body system of field and matter. Using the Keldysh technique for 
the photon Green's function (GF) \protect\cite{Ke65} and neglecting effects of 
spatial dispersion (SD), a formula for the intensity emitted by a steadily excited 
semiconductor into the surrounding vacuum was derived in ref.~\protect\cite{HeKo96}.

In this article this result will be generalized in the following 
fundamentally important aspects: (a) Rigorous results for the emission will 
be given in terms of the exactly defined polarization function of an 
isotropic but otherwise arbitrary medium as an energetically open and 
spatially inhomogeneous finite system. Hence, this approach applies 
independently of any further specific material properties or models and, 
besides others, also SD is exactly considered. (b) Instead of an (electromagnetic) 
vacuum, an arbitrary nonequilibrium distribution of 
photons is assumed to be present. In this way a situation will be addressed 
where neither the radiation nor the medium are in equilibrium but in 
different steadily excited states arbitrarily far from equilibrium. (c) 
Absorption as measured by a linear transmission-reflection experiment is 
investigated on the same footing and its relation to emission is shown.

\section{General theory}
We assume steady state conditions and start with the field operator of the vector 
potential in Coulomb gauge, which obeys the inhomogeneous wave equation
\begin{equation}
\left[\,\, \Delta - \frac{1}{c^2}\frac{\partial^2}{\partial t^2}\,\, \right] \,\, \hat{\bf A}({\bf r},t)
    = -\mu_0 \,\,[ \,\,\hat{\bf j}({\bf r},t) + {\bf j}_{\rm ext}({\bf r},,t) \,\,]
\label{AGFeq}
\end{equation}
and the equal time commutation rules
\begin{equation}
\left[\hat{{\bf A}}({\bf r},t), \frac{\partial}{\partial t}\,\hat{{\bf A}}({\bf r}'
\!,t)\right]     = \frac{i\hbar c}{\varepsilon_0 }\, \overleftrightarrow{t}(
{\bf r}{-}{{\bf r}'}),\label{AEcommut}
\end{equation}
where $\overleftrightarrow{t}$ is the transverse delta function tensor. 
In eq.~(\ref{AGFeq}), the total transverse current density operator 
is split into the current density operator $\hat{{\bf j}}$ of the medium and an 
externally controlled current density represented by given c-number functions 
${\bf j}_{\rm ext}({\bf r},t)$. 

The Keldysh components of the photon GF
\begin{multline}
   D_{ij}^>({\bf r},{\bf r}'\!,t{-}t')=D_{ji}^<({\bf r}',{\bf r},t'-t)\\
   = \frac{1}{i\hbar}\left[\langle
   \hat A_i({\bf r},t)\hat A_j({\bf r}',t')\rangle-\langle
   \hat A_i({\bf r},t)\rangle\langle \hat
   A_j({\bf r}',t')\rangle\right]\label{8b} \, ,
\end{multline}
describe the field-field fluctuations. Formally solving the Dyson equation 
for these components, one obtains the so-called \emph{optical theorem} after
Fourier transformation with respect to $t-t' \rightarrow \omega$ 
\begin{multline}
D_{ij}^\gtrless({\bf r},{\bf r}',\omega) =  \int \upd^3{\bf r}_1 \upd^3{\bf r}_2\;
\times \\
 D_{ik}^{\rm ret}({\bf r},{\bf r}_1,\omega) P_{kl}^\gtrless({\bf r}_1,{\bf r}_2,
\omega)\,D_{lj}^{\rm adv}({\bf r}_2,{\bf r}',\omega),  \label{8c}
\end{multline}
where $P_{kl}^\gtrless$ are the Keldysh components of the polarization function,  
which is given by the functional derivative (in compact notation: 
$P=\delta j/\delta A$) of the induced (averaged) current density 
${\bf j} = \langle \, \hat{\bf j} \, \rangle$ to the mean field 
${\bf A} = \langle\hat{\bf A}\rangle$.

Throughout this paper, $P=\delta j/\delta A$ is to be taken 
at $A = 0$, \ie, in the linear approximation. This is evident since linear 
absorption is addressed, on the one hand, and since the incoherent emission is 
defined as the one without any incident classical field, \ie, 
${\bf A}({\bf r},t) = 0$, on the other hand.
 
\section{Slab geometry}
A slab of thickness $L$ will be considered, which is infinitely extended and 
homogeneously excited in the transverse $y$-$z$-direction. For notational 
simplicity, TE-polarized light propagating freely in the transverse 
direction is considered. Due to cylindrical symmetry around the $x$-axis, 
the transverse vector potential $\hat{\bf A}({\bf r},t)$ can be chosen in the 
$z$-direction.

\section{Classical field propagation} 
Assuming steady state conditions, the propagation equation 
for the average field, after Fourier transforming 
with respect to $\,(y,z) \rightarrow {\bf q}_\perp\,$ and 
$\,t-t' \rightarrow \omega\,$, in this geometry has the structure 
\begin{eqnarray}
\label{1}
\int \upd x' \, D^{{\rm ret},-1} \, (x,x') \, A(x') 
= - \mu_0 \, j_{\rm ext} (x) \,, && 
\end{eqnarray}
\begin{eqnarray}
\nonumber\
D^{{\rm ret},-1}(x,x') 
= \left( \frac{\partial^2}{\partial x^2} + q^{2}_0 \right) \,
\delta (x-x') - P^{\rm ret} (x, x') \,.
\end{eqnarray}
Here and in what follows, the variables ${\bf q}_\perp$ and $\omega$, which 
enter all equations parametrically only, are omitted where possible. 
The wave vector in vacuum is 
$q_0^2 = [(\omega + i\delta)/c)]^{2} - q_{\perp}^{2}$, and
$D^{{\rm ret},-1}$ is the inverse of the retarded photon GF. 
The retarded polarization function of the medium 
$P^{\rm ret}(x,x', q_\perp , \omega)$ 
is related to the linear susceptibility $ \chi $ according to 
$ P^{\rm ret} (x,x') = - \omega^2 \chi (x,x')/c^2 $.
It reflects the translational invariance in transverse directions and 
enables to exactly include SD. 

For a traditional transmission-reflection experiment, the external source on the 
right-hand side of (\ref{1}) is to be put zero, and instead an 
external wave incoming from left or right is assumed. Consequently, the 
homogeneous propagation equation (\ref{1}) has two linearly independent 
solutions, which give the solution of the reflection-transmission problem 
for incidence from left or right. These solutions are fixed by their 
asymptotics and there is no need to impose any further boundary conditions on 
them (even not Maxwell's), since the polarization of the medium increases 
continuously in the transition region from vacuum to medium and, 
consequently, the solutions of (\ref{1}) evolve continuously, too, from the 
asymptotic ones (compare also the discussion in 
\protect\cite{MaMi73}). Only if one assumes an abrupt switch of the 
polarization at the surface, one has to make sure the continuity of the 
solutions of (\ref{1}) by imposing Maxwell's boundary conditions. 

\section{Transmission and reflection}
In the following it is assumed that up to a negligible error the surface of 
the medium, \ie, the length $L$ of the slab, can be fixed in a way that any 
polarization vanishes outside and, hence, any Keldysh component of the 
polarization function vanishes outside, too. Then the forward propagating 
solution of the homogeneous equation (\ref{1}) has the structure
\begin{eqnarray}
\label{2}
A(x) \, = \, \left\{
\begin{array}{lcl}
e^{i q_0 x} +  r \, e^{-i q_0 x} & \quad \mbox{for} \quad & x < - \frac
{L}{2} \\[0.2cm]
t e^{i q_0 x}  & \mbox{for} & x > \frac{L}{2} \, ,
\end{array}
\right.
\end{eqnarray}
where $\, r,t \,$ are reflection and transmission coefficients for the field 
amplitudes. Due to the symmetry $x \leftrightarrow -x$, $A(-x)$ is a 
solution, too (backward propagating, \ie, incidence from right). The 
solutions inside need not to be specified. Neglecting spatial dispersion, 
\ie, assuming $P^{\rm ret} (x,x') = P^{\rm ret} \cdot \delta (x-x')$,
yields inside $A(x) = a e^{iqx} + b e^{-iqx}$, where the wave vector 
is given by $q^{2} = q^{2}_0 - P^{\rm ret}$. In this case, the 
results of ref. \protect\cite{HeKo96} can be reproduced.

Many attempts have been made to consider SD by tracing back this problem to 
the one of the bulk limit, where due to full spatial homogeneity the 
polarization functions $P ({\bf q},\omega)$ can be handled. Most 
common is the use of bulk (polariton) solutions for the waves (\ref{2}) 
inside. This, however, is not consistent with eq. (\ref{1}) and introduces a 
high degree of arbitrariness into the problem, since additional boundary 
conditions (ABC's) \protect\cite{AgGi66} have to be used. The dielectric 
approximation \protect\cite{MaMi73} uses the bulk susceptibility inside. 
It is free of arbitrariness but thought to conflict with energy conservation 
\protect\cite{RiFlHe07}.
The present author suggested to construct solution (\ref{2}) inside 
as induced by sources in a surface region \protect\cite{He98}. 
This procedure is free of arbitrariness if the surface region is very small and the
sources can be handled as $\delta$~functions. 
This condition is obviously fulfilled in some 
specific materials, where this approach worked well \protect\cite{NeVi05}. 
However, for excitons in semiconductors it was shown by microscopic 
investigations \protect\cite{ScJa01} that all the mentioned proposals fail 
to reproduce spectroscopic details correctly.
 
\section{Poynting's theorem for classical fields}
Now Poynting's theorem will be exactly addressed on the grounds of 
eqs.\ (\ref{1},\ref{2}) only, \ie, without referring to any of the above 
mentioned approaches that consider SD approximatively. Quite generally it 
reads 
\begin{eqnarray}
\label{3}
\frac{\partial}{\partial t} \int U({\bf r},t) \,\upd V \, + \int {\bf S}({\bf 
r},t) \, \upd {\bf f} = - \int W({\bf r},t) \, \upd V \,.
\end{eqnarray}
Here, the change of the field energy $ U $ will not contribute to any of the 
quantities discussed below and, hence, can be omitted. Moreover, Poynting 
vector 
${\bf S} = \frac{1}{\mu_0} ({\bf E} \times {\bf B})$, 
${\bf E} = - \frac{\partial {\bf A}}{\partial t}$, 
${\bf B} = \mbox{curl}\,{\bf A} $, and the absorption 
$ W = {\bf j} \cdot { \bf E}$ have been introduced, where the current 
density defines via ${\bf j} = \frac{\partial {\bf P}}{\partial t}$
the polarization $ {\bf P} = \chi {\bf E}$. 

Integrating over the slab in eq.\ (\ref {3}) and Fourier transforming 
with respect to $ t \rightarrow \omega $, this simplifies to 
\begin{eqnarray}
\label{4}
S\left(\frac{L}{2},\omega \right)-S\left(-\frac{L}{2},\omega 
\right)=-\int\limits_{-\frac{L}{2}}^{\frac{L}{2}} \upd x \, 
W(x,\omega)
\end{eqnarray}
for the geometry investigated here. In eq.\ (\ref{4}), $S(x,\omega)$ 
is the $x$-component of the Poynting vector.\\

\section{Coherent absorption}
At first, Poynting's theorem (\ref{4}) will be addressed for average fields 
(classical light). Assuming a monochromatic wave of frequency $\omega_0$ incident 
in the $(q_0 , {\bf q}_{\perp,0})$-direction, \ie, $A (x, {\bf q}_\perp , \omega) = \frac{1}{2} [A_0 (x, {\bf q}_{\perp , 0} , 
\omega_0) \delta (\omega - \omega_0) 
\delta_{  {\bf q}_\perp ,{\bf q}_{\perp , 0} } +  A_0^* (x, {\bf q}_{\perp 
, 0} , \omega_0) \delta (\omega + \omega_0) 
\delta_{  {\bf q}_\perp ,-{\bf q}_{\perp , 0} }]$,
the static part $\propto \delta(\omega)$ of the energy balance (\ref{4}) 
can be regarded separately. Using this ansatz and (\ref{2}) in (\ref{4}) 
yields after straightforward calculation for each 
$\omega_0 \rightarrow \omega$ and ${\bf q}_{\perp,0}  \rightarrow 
{\bf q}_{\perp} $
\begin{multline}
\label{8}
1-|r|^2 - |t|^2 = a = \\
\frac{i}{2q_0} \int \upd x \upd x' A^{*}(x)\hat{P}(x,x')A(x') .
\end{multline}
On the RHS, due to well-known GF identities \protect\cite{Ke65} 
\begin{eqnarray} 
\label{8a}
\hat{P} =  P^{\rm ret} -  P^{\rm adv} = P^> - P^< \,\, ,
\end{eqnarray}
the coherent absorption ($a>0$) or gain ($a<0$) balances generation $ i P^>(x,x')$ 
and recombination $  i P^<(x,x')$ of excitations in the medium, whereas the LHS 
balances the incoming intensity $(\propto 1)$ into absorption and the sum of the 
reflected and transmitted intensity. The latter may be regarded as the intensity 
re-emitted coherently to the incoming light and is to be contrasted with the 
incoherent (or correlated) emission, which will be addressed now. 

\section{Poynting's theorem for field fluctuations}
The incoherent emission is defined as the one without external sources, \ie, for 
vanishing average fields. In this case, Poynting's theorem is still given by 
eq.\ (\ref{3}), but due to the non-commuting field operators 
($ \hat{\bf E} $ and 
$ \hat{\bf B}$ as well as 
$\hat{\bf j} = \hat{\bf j}_0 - e \hat{\bf A}$ and $\hat{\bf E}$) the 
symmetrized Poynting vector 
$\hat{\bf S} = \frac{1}{2 \mu_0} (\hat{\bf E} \times \hat{\bf B} - \hat{\bf 
B} \times \hat{\bf E})$ and the energy dissipation 
$\hat{W} = \frac{1}{2} (\hat{\bf j} \hat{\bf E} + \hat{\bf E} \hat{\bf j})$, 
respectively, have to be used. Their quantum statistical averages are given by the 
Keldysh components of the photon GF (\ref{8b}). 

For slab geometry and TE polarization, the quantum statistical averages of both the 
Poynting vector $\hat{\bf S}$ and the energy dissipation $\hat{W}$ are defined 
via $(\omega,{\bf q}_\perp)$-integrals (for details of the derivation see, \eg, 
ref. \protect\cite {HeKo96}). In the following, these integrals will be omitted 
for notational simplicity, and their spectrally and directionally resolved 
contributions (both scaled by $1/\hbar \omega$) for slab geometry  denoted by 
$s$ and $w$ (in contrast to coherent absorption $a$), respectively, 
will be considered instead:
\begin{eqnarray}
\label{12}
s(x) = -\frac{1}{\mu_0} \left\{ \frac{\partial}{\partial x'} \, 
\left[ D^{>} (x,x') + D^{<} (x,x') \right] \right\}_{x'=x} 
\, ,
\end{eqnarray}
\begin{multline} 
\label{12a}
w(x) = \frac{1}{2}\int \upd x'[ P^{>}(x,x')D^{<} (x',x)\\ - 
 P^{<}(x,x') D^{>}(x',x) ] .
\end{multline}
$ D^{\gtrless} (x,x', {\bf q}_\perp , \omega)$
and $ P^{\gtrless} (x,x', {\bf q}_\perp , \omega)$ are the Keldysh 
components of the photon GF and of the polarization function, respectively. 
Their parametric dependence on the variables ${\bf q}_\perp , \omega$ 
is, as before, not explicitly written in the equations. Note that both the 
incoherent (or correlated) energy flow $s$ and dissipation $w$ are time-independent 
in the steady state, in contrast to the classical case. The $\omega-$ and 
${\bf q}_\perp $-integrals omitted here indicate merely that field fluctuations of 
all frequencies and directions contribute to them. 

\section{Medium- and vacuum-induced contributions}
The polarization function 
\begin{eqnarray}
\label{13}
P^{\gtrless} (x,x') = P^{\gtrless}_m (x,x')
- i \delta \, \frac{4 \omega}{c^{2}} n^{\gtrless} \, \delta (x-x')
\end{eqnarray}
comprises besides the medium part an infinitely weak ($\delta\rightarrow0$) part 
which describes the vacuum in absence of the medium and ensures the correct vacuum 
limit for the \emph{optical theorem} (\ref{8c}) \protect\cite
{HeKo96}. $ n({\bf q}_\perp , \omega ) = n^< = n^> -1 $ describes 
a given nonequilibrium distribution of photons owing to external preparation as, 
\eg, by an appropriate enclosure (heath bath) or incoherent radiation incident 
from outside. Therefore $n$ in the following will be referred to as the distribution of 
the externally controlled or simply of external photons. As a consequence, the 
\emph{optical theorem} (\ref{8c}) comprises a medium-induced and a vacuum-induced 
contribution according to
\begin{eqnarray}
\label{14}
D^{\gtrless}(x,x') = D^{\gtrless}_{m}(x,x')
+ n^{\gtrless}\hat{D}_0 (x,x') \, ,
\end{eqnarray}
where 
\begin{multline}
\label{14a}
D^{\gtrless}_m (x,x')=\\
\int \upd \bar x \, \upd \bar x'
D^{\rm ret}(x,\bar x)P^{\gtrless}_m(\bar x,\bar x')D^{\rm adv}(\bar x' , x')
\end{multline} 
and
\begin{eqnarray}
\label{14b}
\hat{D}_0 (x,x') = - i \delta \, \frac{4 \omega}{c^{2}} \int \upd \bar x \, D^{\rm 
ret} (x,\bar x) \,
D^{\rm adv} (\bar x , x') \,\, .
\end{eqnarray} 
Correspondingly, the 
spectral function $ \hat{D} \equiv  D^{\rm ret} - D^{\rm adv} = D^> - D^< $ 
decomposes into a medium-induced and a vacuum-induced contribution, too, 
\begin{eqnarray}
\label{14c}
 \hat{D}(x,x') = \hat{D}_m (x,x') + \hat{D}_0 (x,x') \, , 
\end{eqnarray} 
where 
\begin{multline}
\label{14d}
\hat{D}_m (x,x')= \\
\int \upd \bar x \, \upd \bar x'
D^{\rm ret}(x,\bar x) \hat{P}_m(\bar x,\bar x')D^{\rm adv}(\bar x' , x') \, .
\end{multline} 
Note that $\hat{D}_0$ is to be contrasted with the spectral function 
of the pure vacuum $\hat D_{\rm vac}(x,x') = \cos [q_0(x-x')]/2iq_0 $. Also, it 
contains an improper integral diverging as $1/\delta$, which is compensated by the 
prefactor $\delta \to 0$. Therefore, evaluating the $\bar x$-integral for 
$\hat{D}_0 $, it is  sufficient to regard $|\bar x| > L/2$, 
where $\,\, D^{\rm adv} (\bar x,x') = D^{\rm ret}(x',\bar x)^{*} \,\,$ 
is given in terms of solution (\ref{2}) as
\begin{eqnarray}
\label{17}
D^{\rm ret} (x,x') = \frac{ \Theta (x-x') A(x) A(-x')}{2iq_0 t} \,
+ ... x \leftrightarrow x' ...    
\end{eqnarray}
Inserting (\ref{17}) in (\ref{14b}) yields
\begin{eqnarray}
\label{18}
\hat{D}_0 (x,x')=\frac{1}{2iq_0} 
[A(x) A^{*}(x')+A(-x)A^{*}(-x')] .
\end{eqnarray} 
Using (\ref{13}) - (\ref{18}) in (\ref{12a}) for $\omega>0$ yields for the energy 
dissipation
\begin{eqnarray}
\label{19}  
w = \int \upd x \, w(x) = \frac{i}{q_0} [n^< \mathfrak{P}^> - n^>  \mathfrak{P}^< ], 
\end{eqnarray}
where the global generation/recombination 
$\mathfrak{P}^{\gtrless}(\omega, {\bf q}_\perp)$ are defined as
\begin{eqnarray}
\label{19a}
\mathfrak{P}^{\gtrless} = \int \upd x \, \upd x' A^{*}(x)  \, P^{\gtrless}_m (x,x') \, A (x') \,\, . 
\end{eqnarray} 
It is noteworthy that, inserting (\ref{14}) in (\ref{12a}), all the contributions 
induced by the medium according to (\ref{14a}) cancel exactly and, thus, 
contribute neither to the (incoherent) energy dissipation nor to the emission.  

Equation (\ref{19}) balances globally optical excitation $i\mathfrak{P}^>$ 
of the medium accompanied by absorption of external photons ($\propto n$) and 
recombination $i\mathfrak{P}^<$ accompanied by emission of photons ($\propto [1+n]$,
\ie, spontaneous and externally stimulated emission). 

\section{Nonequilibrium photon distribution}
As usual \protect\cite{Ke65}, a distribution $ b(\omega, {\bf q}_\perp) \equiv  
b^< = b^> - 1 $ will be attributed to the global generation/recombination 
$\mathfrak{P}^{\gtrless}(\omega, {\bf q}_\perp)$  in (\ref {19a}) by definition
\begin{eqnarray}
\label{20}
\mathfrak{P}^{\gtrless}
= b^{\gtrless} \hat{\mathfrak{P}}\,\, , 
\end{eqnarray} 
where $ i \hat{\mathfrak{P}} = i ( \mathfrak{P}^> - \mathfrak{P}^< ) = 2 q_0 a $  
is directly related to the coherent absorption $ a $ in eq.\ (\ref{8}). In contrast 
to $n$, the occupation $b$ characterizes globally the distribution of the 
medium-induced (transverse) optical excitations (\eg, polaritons in semiconductors) 
over the absorption/gain spectrum.

Using (\ref{20}) in (\ref{14}), \eg, for $ (x,x')>L/2$, the field fluctuations 
take the explicit form 
\begin{eqnarray}
\label{20a}
D^{\gtrless}(x,x') = b^{\gtrless}  \hat{D}_{m}(x,x')
+ n^{\gtrless}\hat{D}_0 (x,x') \, ,
\end{eqnarray} 
where the medium-induced part of the spectral function in contrast to $\hat{D}_0$ 
is related to the classical absorption $a$ as 
\begin{eqnarray}
\label{20b}
2iq_0 \hat{D}_m (x,x') = a e^{iq_0(x-x')} \, .
\end{eqnarray}

Now Poynting's theorem (\ref{4}) provides the incoherent 
energy flow $ S(L/2) = -S(-L/2) $ from the energy dissipation 
(\ref{19}): $ 2s(L/2) = - w $. Then from (\ref{4}) and (\ref{19}), for any 
frequency $\, \omega > 0 \,$ and propagation direction 
$\,\, {\bf q} = (q_0 , {\bf q}_\perp) \,$, we obtain for the spectrally and 
directionally resolved energy flow 
\begin{eqnarray}
\label{21}  
s = -\frac{i}{2q_0} (n^< \mathfrak{P}^> - n^> \mathfrak{P}^<) 
 = ( b - n ) a \, .
\end{eqnarray}
In eq.\ (\ref{21}), $ -n a $ describes absorption ($a>0$) or emission 
($a<0 $) as the response of the medium to (and stimulated by) the given 
nonequilibrium distribution of external photons $n$. Putting $ n = 0 $,
the emission $ ba $ is the response of the medium to vacuum fluctuations. It is to 
be regarded as spontaneous emission with respect to external photons, but it is 
stimulated by $b$. It is not directly related to the coherent absorption. Instead, 
the recombination $i\mathfrak{P}^<$, and so $b$, has to be calculated from the 
particle GF's of the interacting many-body system of field and matter on just the 
same footing as has been carried out here for the photon GF. 

Assuming $ s = 0 $ , \ie, vanishing energy flow between the
medium and its surrounding, one arrives at $ b = n $, which generalizes 
Kirchhoff's law to nonequilibrium, and $b$ develops towards a Bose function 
$\,b (\omega) = \left( {\rm exp} \left[\beta \hbar \omega \right] - 1 
\right)^{-1}\,$ for thermodynamic equilibrium. 

If the externally given incoherent radiation field incident from outside is 
strong, \ie, for $ n\gg  b$, measuring the energy flow $s$ would provide the same 
information as a classical (coherent) reflection-transmission experiment, namely 
the absorption $a=s/n$. 
 
In the following, the opposite case $ b \gg  n \rightarrow 0$, \ie, emission 
$ s = b a $ into the surrounding vacuum is considered. Then the distribution $b$ is 
accessible to direct observation in experiments measuring simultaneously the 
(incoherent) emission and the linear coherent absorption, \ie, transmission and 
reflection. It generalizes Planck's formula for the black body radiation to the 
nonequilibrium radiation of an excited medium in the steady state. 

\section{Quasiequilibrium}
Of particular interest are those steadily excited states which 
can be regarded as quasi-equilibrium states. As such,
\eg, for semiconductors, exciton gases generated at 
low up to moderate excitation and light-emitting diodes 
working at high excitation are to be mentioned. 
For quasi-equilibrium, due to the Kubo-Martin-Schwinger 
condition \protect\cite {KMS57}, the distribution $b(\omega, {\bf q}_\perp)$ 
develops into a Bose distribution $\,b (\omega) = \left( {\rm exp} 
\left[\beta(\hbar \omega - \mu)\right] - 1 \right)^{-1}\,$, being independent 
of $ {\bf q}_\perp $. The chemical potential $ \mu $ 
starts at $ \mu = 0 $ for complete thermal equilibrium and characterizes 
the degree of excitation beyond the thermal one for $ \mu > 0 $. 
Then the crossover from absorption ($a>0$) to gain ($a<0$) appears 
independently of ${\bf q}_\perp$ at $ \hbar \omega = \mu $, 
where the singularity in $b$ is compensated by the zero in $a$. 
Hence, expanding $b^{-1}$ and $a$ at $\hbar \omega = \mu $ 
yields that at crossover the emission stays 
finite and is given by the slope of the absorption according to
$ s(\mu,{\bf q}_\perp) = k_B T \left\{ \partial 
a(\omega,{\bf q}_\perp)/\partial \omega \right\}_{\hbar \omega = \mu}$.
Since both $\, a = 1 - |r|^2 - |t|^2 \,$ and $b$ switch their signs, the 
emission $ s $ stays positive as it should be in the whole frequency region. 
Measuring $ s $ and $ \, a = 1 - |r|^2 - |t|^2 \,$ would enable 
to check whether quasiequilibrium is realized. If so, the chemical 
potential $\mu$ is fixed through the crossover point and, after that, the 
temperature and excitation density can be obtained directly from 
experimental data. 

\section{Low temperature}
For $T \to 0$, the Bose function degenerates to a step function and the emission 
$s(\omega,{\bf q}_\perp) \to - \Theta (\mu - \hbar \omega) a(\omega,{\bf q}_\perp)$ 
vanishes completely in the absorption region $  \hbar \omega > \mu $ and reflects exactly the gain $-a$ in the gain region $  \hbar \omega < \mu $. 

Concerning low temperatures, however, the theory given above needs to be 
supplemented if effects of quantum condensation occur. As such, the crossover 
from Bose-Einstein condensation of excitons at moderate excitation to the one of 
Cooper-like electron-hole pairs at high excitation has been addressed \protect\cite
{Noz}. Its consequences for emission and absorption spectra will be discussed in 
detail in a forthcoming paper. However, some basic features should be commented on 
here. If quantum condensation occurs, an anomalous contribution to 
$P^{\gtrless}_m \rightarrow P^{\gtrless}_m + P_{\rm cond} \delta_{{\bf q}_\perp 
,0} \delta(x-x') \delta(\hbar \omega - \mu)$ will appear in addition 
to the normal generation $ i P^>_m$ and recombination $ i P^<_m$ 
considered above. The Kronecker $\delta_{{\bf q}_\perp , 0}$ indicates that 
this term is to be excluded from the ${\bf q}_{\perp}$ integrals. 
The strength $ P_{\rm cond} $ is determined by the fraction of 
quasiparticles in the condensate. Since it appears identically in both 
the generation $ iP^>_m $ and the recombination $ iP^<_m $, its influence 
cancels in the classical absorption $ a $ according to (\ref{8}) and (\ref
{8a}). Consequently, those effects will not appear directly in classical 
absorption experiments, where at best they show up as smooth changes in the 
spectral shape of the absorption $a$. The same applies to the absorption ($a>0$) 
or emission ($a<0 $) as the response of the medium to (and stimulated by) the 
externally given nonequilibrium distribution of photons $n$ in (\ref{21}). 

However, an additional sharp peak in the emission 
\begin{eqnarray}
b a = \frac{1}{2 q_0} i  \mathfrak{P}^< 
\end{eqnarray}
at $ \hbar \omega = \mu $, whose 
strength is $\propto P_{\rm cond} \int \upd x \, \Theta (L/2 - |x|) | A(x) |^2 $, 
would give evidence for a condensate, since the normal part of the emission
just at this frequency tends towards zero. 

\section{Conclusion}
Using a quantum-kinetic many-body approach, exact results have been presented for 
the interacting system of field and matter in a specified geometry. The 
spectral function of photons splits up into a vacuum-induced and a medium-induced
contribution (see eqs.\ (\ref{14b}-\ref{14d}), for which the explicit expressions 
(\ref{18}) and (\ref{20b}), respectively, have been obtained. It is noteworthy 
that both kinds of states are globally (\ie , inside {\it and} outside) defined and 
their split does not correspond to a spatial seperation of the inside from the outside. 
There are, of course, many different options to split up the spectral function of 
photons, \eg, into the one of the pure vacuum $\hat D_{\rm vac}$ and the remaining. 
But only the decomposition according to eqs.\ (\ref{14b}-\ref{14d}) 
yields an exact cancellation of the 
medium-induced contribution in the balance (\ref{12a}) and so the physically clear 
and simple structure of the emission (\ref{21}). 

Also, according to eq.\ (\ref{20a}), the field fluctuations split up into a 
vacuum-induced and a medium-induced contribution whose strengths are given
by the distributions $n$ and $b$, respectively. 

The distribution $n$ describes the population of the vacuum-induced states and is 
externally given through preparation of the surrounding, \eg, either as a heath 
bath or by incoherent radiation incident from outside. Measuring the net energy 
flow $ -na$ induced by an incoherent radiation $n$ incident from the outside provides 
the absorption $a$ just as a classical (coherent) transmission-reflection 
experiment. 

The distribution $b$ describes the population of the medium-induced states. It 
is fixed by the steady-state excitation conditions of the medium and characterizes 
its globally defined transverse optical excitations as, \eg, excitonic polaritons 
in semiconductors. For a medium in thermal equilibrium, $b$ tends towards a Bose 
function, to which a chemical potential can be attributed in the case of 
quasi-equilibrium. Thus, in a sense, real photons, \ie \ after their interaction 
with the fermions of the medium is exactly considered, behave statistically like 
ideal bosons. \Eg, for semiconductors, this applies likewise to optical 
excitations below (excitonic polaritons) and above the fundamental gap, whereas 
neither excitons nor ionized electron-hole pairs can be regarded as (ideal) bosons.
For $n=0$, \ie, without any preparation of the surrounding of the medium, emission 
is governed by the distribution $b$. It generalizes the Planck distribution for 
ideal photons in thermodynamic equilibrium to interacting ones in nonequilibrium. 

Our results prove that even lasing can be regarded as quasi-thermal 
emission. This has already been demonstrated in ref.~\protect\cite{HeKo96} 
neglecting spatial dispersion and is now exactly confirmed. All the 
novel features of this light result exclusively from an extremely strong 
renormalization of the globally defined coherent absorption $a$ in the gain region 
due to high compensation of output losses there.

Emission out of a quantum condensate appears as an additional sharp peak at the 
crossover point, whereas no significant structures are expected in the coherent 
absorption.

As a challenge for both experimentalists and theorists, the nonequilibrium 
distribution $b$ can be observed directly measuring emission and absorption 
simultaneously \protect\cite{Se07} or has to be computed, respectively, from the global
recombination providing reasonable approximations for the polarization function.

It may appear counterintuitive and surprising that the medium-induced contribution $\hat D_m$ exactly cancels out in the absorption and emission, and the question arises whether this remains true if one moves away from the restrictions to slab geometry and the steady state, or further, which form the energy flow law takes without them. 
The geometry-independence of Kirchhoff's and Planck's laws are bought with severe restrictions in other domains, namely to full thermal equilibrium and noninteracting photons. 
It can hardly be expected to obtain a radiation law which is universal in all these aspects and at the same time more detailed than just the condition of energy conservation, since quantities like absorption or distribution lose their meaning in a  temporally inhomogeneous case. 
Moreover, in nonequilibrium, energy flow in both the medium and its environment have to be evaluated separately for the considered geometry, and the less symmetries the geometry provides, the more difficult this will be. 
Taking this into account, it appears all the more surprising that for the present case of steady-state slab geometry the evaluation can be taken such a long way.
What we know today is at least that the splitting of the spectral function in the presented form is a completely universal property of the photon GF \cite{He08} and will thus always affect Poynting's theorem. Analyzing the latter for a non-steady state, we suppose one will find the interplay of $\hat D_m$ and $\hat D_0$ differ and worth examining closer.

\begin{acknowledgments}
The author wishes to thank the {\it Deutsche Forschungsgemeinschaft} for
support through {\it Sonder\-forsch\-ungs\-be\-reich~652}. Stimulating discussions 
with the members of the semiconductor theory group at the University of Rostock 
and with {\it H. Stolz} and {\it W. Vogel} are gratefully acknowledged.
\end{acknowledgments}



\begin{thebibliography}{10}
\bibitem{RiFlHe07}
Making use of the results of this work, one can show that this is not the case,
see F.~Richter, M.~Florian and K.~Henneberger, Europhys. Lett. {\bf 81}, 67005 (2008).
\end{thebibliography}
\end{document}